\documentclass[11pt,a4paper]{article}

\usepackage{graphicx}
\usepackage{txfonts}

\usepackage{hyperref}

\usepackage{xcolor}
\usepackage{ulem}

\begin{document}

\title{THE LEINSTER-COBBOLD DIVERSITY INDEX AS A CRITERION FOR SUB-CLUSTERING\thanks{Accepted for publication in  \href{https://www.printspublications.com/journal/iapqr-transactions\#journal}{IAPQR Transactions.}}}

\author{HUGO CHAMBON\footnote{E-mail: hugo.chambon@univ-grenoble-alpes.fr} \and DIDIER FRAIX-BURNET \\ \\
Univ. Grenoble-Alpes, CNRS, IPAG, Grenoble (France)}

\date{May 2025}
	
	\maketitle
	
	\begin{abstract}
An automatic procedure to perform
sub-clustering on large samples is presented. At each iteration, the
most diverse cluster is sub-clustered, and the global diversity of the
new classification is compared to the previous one. The process stops if
no improvement is found. The key to our procedure is the use of a
quantitative measure of diversity, called the Leinster-Cobbold index,
that takes into account the similarity between clusters. While this
procedure has been successfully applied on a large sample of spectra of
galaxies, we illustrate its efficiency with two examples in this paper.
\end{abstract}

\textbf{Keywords} Astrostatistics, Unsupervised
classification, Sub-clustering, Diversity measurement

\section{INTRODUCTION}

Unsupervised classification (or clustering) is the art of finding in a
data set structures that can be identified as clusters. There are many
criteria that can provide the optimum number of clusters, but this
optimum depends on the sample under consideration. Actually, the number
of clusters is determined by the separation between the structures and
the ability of the clustering algorithm to discriminate them. If some
clusters are close to each other with respect to the others in the
sample, there is a good chance that they will remain unnoticed. As a
consequence, it might be judicious to increase the resolution of the
classification by performing a clustering of clusters individually, that
is called sub-clustering.

This is a well known practice in biology and is easily understood in the
frame of a hierarchical classification such as the Tree of Life.
However, this also occurs for large samples of high-dimension in other
domains. In astrophysics, Fraix-Burnet et al. ({[}1{]}, 2021) have found
that three steps of sub-clustering were necessary for a sample of 702
248 spectra of galaxies. In the first two steps, the optimum number of
clusters was found to be only 3. It appears that one or two of these
clusters are relatively very small, with properties very different from
the rest of the sample. In the third step, the optimum was still 3
clusters, but they were big enough to deserve a sub-clustering, yielding
respectively 50, 25 and 10 clusters.

In high-dimension problems, dimensionality reduction is a necessity.
Sub-clustering finds another justification in the fact the latent
variables found to discriminate broad categories are not necessarily the
same as those required to detect finer structures within them.

In most cases, a human inspection is necessary to decide whether a
sub-clustering can be useful, and when to stop the process. For
instance, Dubois et al. ({[}2{]}, 2024) made the arbitrary choice of
sub-clustering all clusters having more than 5\% of the sample size, and
only once. To our knowledge, most statistical techniques providing a
more objective criterion requires all sub-clusters to be available in
order to compare the full classifications. Among these techniques, the
clustering tree is popular (e.g. Zappia and Oshlack, {[}3{]}, 2018).
However, in the case of large samples, avoiding useless sub-clustering
could save much computing time.

A priori selection of clusters that could benefit a sub-clustering can
be performed using for instance the silhouette score that measures a
level of heterogeneity. But this is not a hard criterion, since ``higher
heterogeneity'' is indicated by ``a lower median silhouette score or
large variations in the score distribution of a cluster'' (Leary et al,
{[}4{]}, 2023). It certainly requires human examination and decision,
and thus seems difficult to automatize.

In this work, we propose to use a quantitative indicator of the
diversity contained in a cluster to decide beforehand to sub-cluster or
not, and compare the global diversity of the new classification with
respect to the previous one to decide whether to continue. In this
paper, classification is taken as the ensemble of all clusters and
sub-clusters. For this purpose, we use the Leinster-Cobbold measure of
diversity (Leinster, and Cobbold, {[}5{]}, 2012) which has been
developed for biodiversity but can be applied more broadly as a measure
of information.

This paper is organized as follows. The Leinster-Cobbold diversity index
is presented in Sect. 2, followed by our procedure for the unsupervised
classification of spectra of galaxies. In Sect. 3, we demonstrate the
relevance of the Leinster-Cobbold diversity measure in astrophysics on a
real example with a clearcut physical interpretation. Our full procedure
for sub-clustering is illustrated in Sect. 4 with a large sample of
spectra of galaxies which is the part of a more complete study that will
be published in an astrophysical journal. A short discussion is given in
Sect. 5 as conclusion.

In all the examples given in this paper, we use the clustering algorithm
called Fisher-EM (Bouveyron and Brunet, {[}6{]}, 2012). It is a Mixture
Gaussian Model in a discriminative latent subspace available as a
package in R. It has been used and described in many papers in
astrophysics. The readers may in particular refer to Fraix-Burnet et al.
({[}1{]}, 2021), Dubois et al. ({[}2{]}, 2024), and Chambon and
Fraix-Burnet (in preparation), in the context of the present work.

\section{METHODS}

Our strategy is to use a quantitative measure of the diversity in an
automatic and iterative procedure. Given a classification, 1) the
diversity of each individual cluster is computed (Sect. 2.2), 2)
sub-clustering is performed on the cluster showing the higher diversity,
3) the global diversity of the new classification is computed (Sect.
2.3). If the global diversity is increased, we reiterate step 1 with the
new classification. If this diversity does not improve, the procedure is
stopped and the previous classification is kept.

In this procedure, the same family of diversity measures is employed,
but computed in two different contexts, either for individual clusters,
or for the global classification.

\subsection{The Leinster-Cobbold diversity index}

Diversity is a measure of variety often used in ecology and deeply
linked with information theory. The well-known Shannon entropy:

\begin{equation}
H\left( \overrightarrow{p} \right) = \sum_{i = 1}^{n}p_{i}\log\left( \frac{1}{p_{i}} \right)
\end{equation}

where \(p_{i}\) is the relative abundance of the $i$th species (or
clusters) among the $n$ species in the community (or sample), is
not well adapted to evaluate the diversity since it is not an effective
number, i.e. not linear with respect to the number of distinct species.
The Hill numbers (Hill, {[}7{]}, 1973) correct this drawback and are
defined as:

\begin{equation}
	D_{q}\left( \overrightarrow{p} \right) = M_{1 - q}\left( \overrightarrow{p},\frac{1}{\overrightarrow{p}} \right)
\end{equation}

where:

\begin{equation}
	M_{t}\left( \overrightarrow{p},\overrightarrow{x} \right) = \left( \sum_{\displaystyle i \in supp\left( \overrightarrow{p} \right)}^{} p_{i}x_{i}^{t} \right)^{\displaystyle\frac{1}{t}}
	\end{equation}

is the generalized mean, and $q$ the order of the Hill number. The
different orders give different perspectives on the notion of diversity:
\(D_{0}\left( \overrightarrow{p} \right)\) corresponds to the number of
clusters and is called species richness;
\(D_{1}\left( \overrightarrow{p} \right)\) is the exponential of the
Shannon entropy;
$D_{2}\left( \overrightarrow{p} \right) = \displaystyle 1/\sum_{i = 1}^{n}p_{i}^{2}$
is the expected number of trials to obtain a pair of individuals of the
same cluster, and is called the inverse Simpson concentration; and
$D_{\infty}\left( \overrightarrow{p} \right) =  1/\max_{\displaystyle i \in supp\left( \overrightarrow{p} \right)}\left( p_{i} \right)$
indicates how much a single cluster dominates the community.

The definition of the Hill numbers relies on the relative abundance of
entirely dissimilar species. To include possible similarity between
species, Leinster and Cobbold ({[}5{]}, 2012) proposed a modified index:

\begin{equation}
	D_{q}^{Z}\left( \overrightarrow{p} \right) = M_{1 - q}\left( \overrightarrow{p},\frac{1}{\overrightarrow{(Zp)}} \right)
	\end{equation}

where:

\begin{equation}
	(Zp)_{i} = \sum_{j = 1}^{n}Z_{ij}p_{j}
	\end{equation}
	
is called the ordariness of the cluster $i$, and $Z$ is the
similarity matrix between species. In this paper, we use the general
definition of the similarity chosen by Leinster and Cobbold
({[}5{]},2012)~:

\begin{equation}
	Z_{\left\{ ij \right\}} = \exp\left( - {ud}_{ij} \right)
	\end{equation}
	
where \(d_{ij}\) is a distance between clusters $i$ and $j$,
and $u$ is a constant.

The function \(D_{q}^{Z}\left( \overrightarrow{p} \right)\) as a
function of \(q\) is called the diversity profile and is used to compare
the diversity of several classifications. In practice, the four orders
\(q = 0,1,2,\infty\) are sufficient to quantify the diversity and are
relatively easy to compute (Leinster and Cobbold, {[}5{]}, 2012).

\subsection{Deciding on sub-clustering for spectra of galaxies}

The first step in our procedure is to sub-cluster one of the clusters,
logically the one that has the highest internal diversity. Leinster and
Cobbold ({[}5{]},2012)~show that it is possible to estimate
\(D_{q \geq 2}^{Z}\left( \overrightarrow{p} \right)\) without a
classification. For instance,
\(D_{q = 2}^{Z}\left( \overrightarrow{p} \right)\) can be computed by
sampling pairs of individuals at random and calculating the mean of the
similarities of the pairs (see also Leinster, {[}8{]}, 2020). It is thus
possible to compute the intra-cluster diversity for a single cluster.
Because the lower orders characterize the rarest species, aim of the
sub-clustering, we chose to use only the order 2 of the intra-cluster
diversities, to rank the clusters.

The distance \(d_{ij}\) in Eq. (6) must be chosen according to our data.
Spectra of galaxies are high-dimensional vectors of monochromatic fluxes
within the wavelength range of the instrument. Their length goes from
several hundreds to several thousands. The fractional \(L_{p}\) norm
appears to be better suited for the high-dimension, since it lowers the
importance of non-Gaussian noise, and it counters the distance
concentration phenomena when data are normalized (François et al.,
{[}9{]}, 2007). Using a Chebyshev lower bound method (Kában, {[}10{]},
2012), we found that, in our case, \(d_{ij} = L_{p}\) with \(p = 0.8\)
is a better choice than the \(L_{p}\) norm of order 1 (Manhattan) or 2
(Euclidean).

The clustering algorithm used in our studies is a Gaussian Mixture Model
in a latent discriminant subspace called Fisher-EM (Bouveyron and
Brunet, {[}6{]}, 2012). Each sub-clustering yields a different subspace,
revealing new latent features. To compare all clusters and sub-clusters
at a given step, the computation of intra-cluster diversity (or
equivalently \(d_{ij}\)) is performed in the concatenated latent
subspaces of all the previous sub-clustering steps. The individual
spectra are thus projected in this new subspace.

The constant \(u\) in Eq. (6) is chosen to be:

\begin{equation}
	u = u_{i} = \frac{1 - (Zp)_{i}}{L}
	\end{equation}

where $L$ is the number of variables, that is the dimension of the
(concatenated) latent discriminant subspaces, and used to normalize the
distance. The numerator introduces a correction to penalize the relative
ordariness of the cluster $i$ with respect to the other clusters.

\subsection{Stopping criterion}

Once the cluster with the highest intra-cluster diversity is
sub-clustered, the global diversity of the new classification is
computed using the new sub-clusters and compared to that of the previous
classification. The distance \(d_{ij}\) is computed in the concatenated
latent subspaces of the various clusters and sub-clusters (see above).
We here choose:

\begin{equation} 
	u = \frac{1}{L}.
\end{equation}

The orders \(q = 0,1\) and 2 can now be computed and the global
diversity compared between the new and previous classifications.
\(D_{\infty}\left( \overrightarrow{p} \right)\) is not considered since
the dominance of a single cluster is not expected to change much by
sub-clustering only one cluster. The species richness
\(D_{0}^{Z}\left( \overrightarrow{p} \right)\) always increases with
sub-clustering since it is related to the number of clusters, but the
amount of its increase gives a first estimation of the gain in
diversity. On the contrary, the ratio
$D_{0}^{Z}\left( \overrightarrow{p} \right)$/\textit{(number of
clusters)} may decrease and provides another estimation of the gain in
diversity. \(D_{2}^{Z}\left( \overrightarrow{p} \right)\) can be
somewhat problematic in some cases because this order characterizes
something intermediate between rare species and dominant ones (Leinster
and Cobbold, {[}6{]}, 2012).

As a consequence, our stopping criterion is divided into two stages.

First, if \(D_{0}^{Z}\left( \overrightarrow{p} \right)\) increases by
more than 30\%, the sub-clustering is considered as improving the
classification if \(D_{1}^{Z}\left( \overrightarrow{p} \right)\)
increases as well. If \(D_{1}^{Z}\left( \overrightarrow{p} \right)\)
decreases, we stop the sub-clustering procedure and keep the previous
classification as definitive.

Second, in the case that \(D_{0}^{Z}\left( \overrightarrow{p} \right)\)
increases by less than 30\%, if any of the three quantities
$D_{0}^{Z}\left( \overrightarrow{p} \right)$/\textit{(number of
clusters)}, \(D_{1}^{Z}\left( \overrightarrow{p} \right)\), and
\(D_{2}^{Z}\left( \overrightarrow{p} \right)\), decreases, the
sub-clustering is considered as not interesting and the previous
classification is kept as definitive.

If the tests show that the sub-clustering improves the diversity, then
the new classification is kept and the procedure is repeated with the
selection of the most diverse cluster (Sect. 2.2).

\section{INTRA-CLUSTER DIVERSITY AND SUB-CLUSTERING: AN ASTROPHYSICAL
CONFIRMATION}

\begin{figure}[h]
	\begin{center}
		\includegraphics[width=0.8\linewidth]{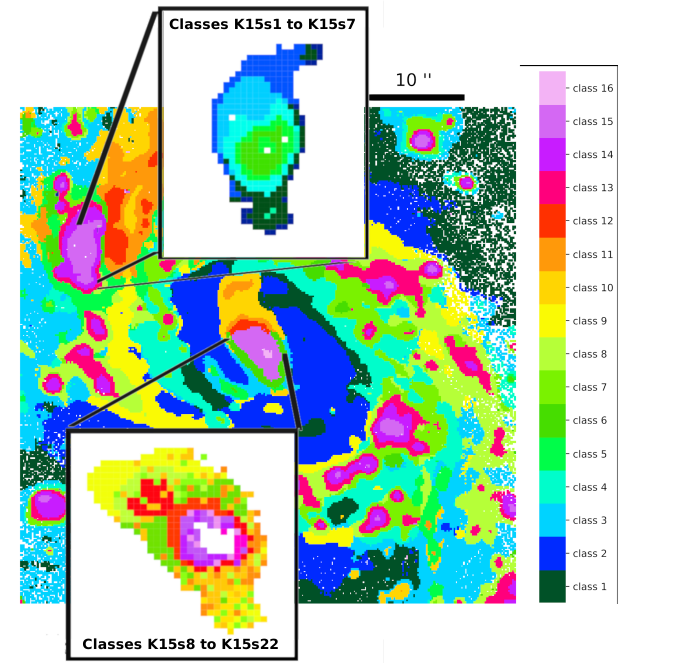}
	\end{center}
	\caption{Subclassification of class 15. The background image shows the initial classification of the hyperspectral image into 16 clusters (right scale bar). The inlets show the results of the sub-clustering of class 15 into 22 sub-clusters. The colors of the sub-classes in the inlets do not correspond to the right scale bar.}
\end{figure}

The integral field spectroscopy is a technique in which each element
(called spaxel) of an image is a spectrum. In this example, we use the
clustering of the spectra included in the hyperspectral image of the
prototypical Seyfert 2 galaxy NGC 1068 presented in Chambon and
Fraix-Burnet ({[}11{]}, 2024). This galaxy shows a powerful starburst
activity concentrated in a prominent starburst ring, together with a jet
that interacts with the gas in the disc of the galaxy. The gas is also
illuminated by the radiation from the Active Galactic Nucleus at the
origin of the jet. This is thus quite a complex objects with many
regions showing various spectral characteristics.

The first clustering step identified 16 clusters with specific
spectroscopy characteristics. This means that all the different regions
in the image can be regrouped into 16 spectral categories, independently
of their position in the image. Chambon and Fraix-Burnet ({[}11{]},
2024) noticed that the cluster 15 (called class 15 in their paper) is
split into several zones, and more particularly into two distinct and
large regions in the galaxy that are not supposed to be illuminated by
the same physical mechanisms. Their spectra were gathered together
because they share a common property, distinct from all the other
spectra in the hyperspectral image: they have a higher continuum level.
The sub-clustering of this class 15 yielded 22 sub-clusters that clearly
distinguish the nucleus from other regions (Fig. 1).

\begin{figure}[h]
	\begin{center}
		\includegraphics[width=\linewidth]{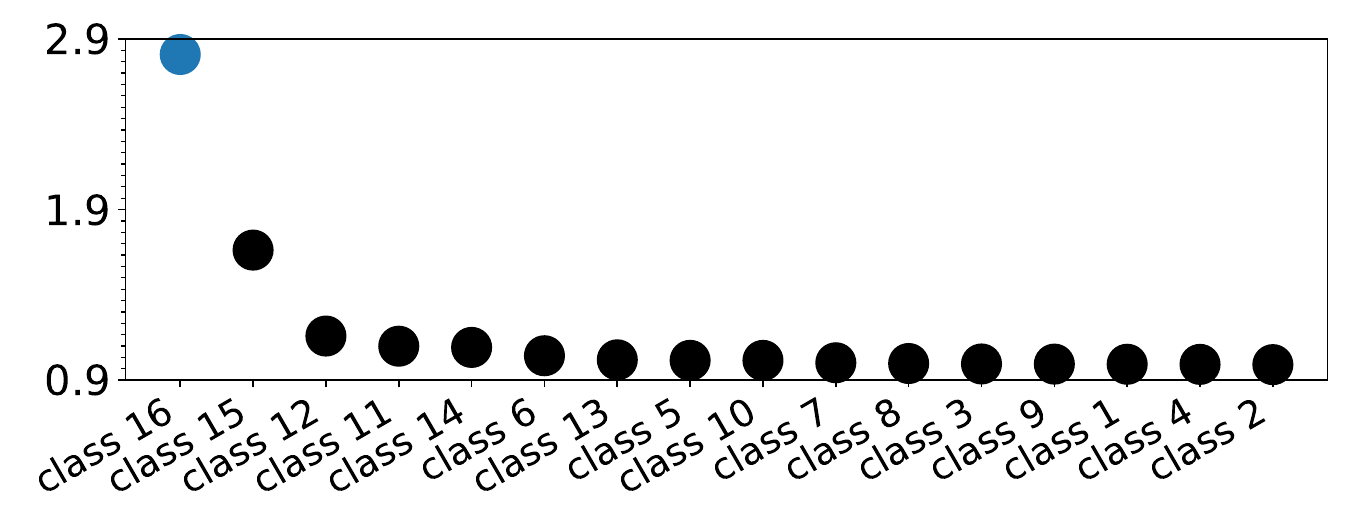}
	\end{center}
	\caption{Estimation of $D_{2}^{Z}\left( \overrightarrow{p} \right)$ for individual classes in NGC 1068. Apart from class 16 which is composed of ill-calibrated spectra, class 15 has a diversity much higher that all the other classes, justifying its sub-clustering.}
\end{figure}

The diversity measure proposed in this paper were not available to
Chambon and Fraix-Burnet ({[}11{]}, 2024). It appears that the
intra-cluster diversity of each cluster (Fig. 2) shows that clusters 16
and 15 are the two most diverse ones. Cluster 16 gathers spectra from
the very centre of the galaxy that are clearly affected by poor
calibration and alignment. This cluster can be ignored in the present
discussion. The diversity of class 15 is much higher than that of the
others, justifying its sub-clustering. This example provides a physical
confirmation of the relevance of the diversity measure of Leinster and
Cobbold ({[}5{]}, 2012) in astrophysics.

\section{APPLICATION OF THE PROPOSED PROCEDURE ON A HIGH-DIMENSION
SAMPLE}

\begin{figure}
	\includegraphics[width=\linewidth]{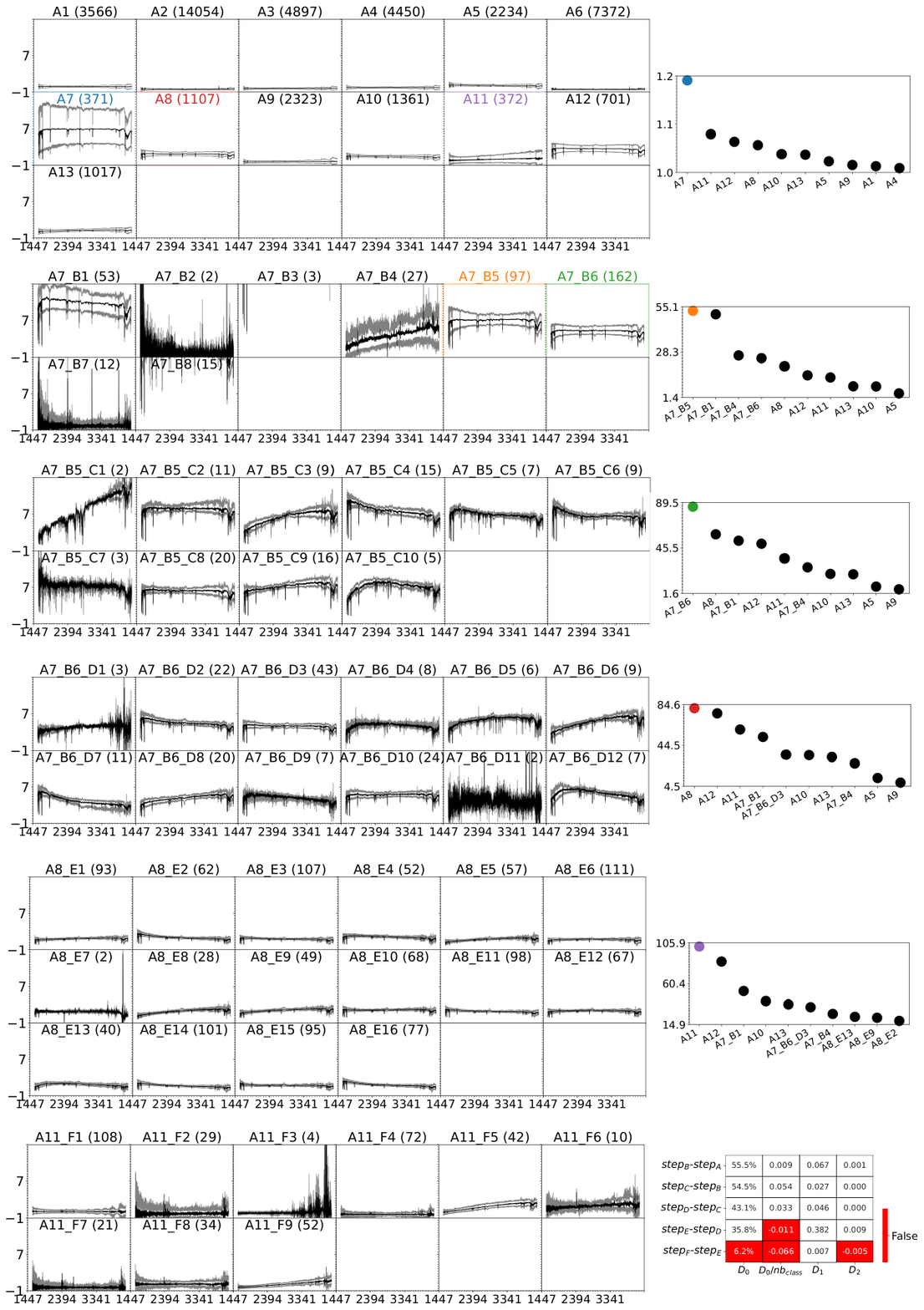}
	\caption{Illustration of our procedure applied to an ensemble of 43826 spectra of galaxies. Each step is identified by a letter in colour. On the left side, the mean spectrum (black), and 10 and 90\% quantiles (gray) for each cluster are shown. On the right, the intra-cluster diversity is plotted for each step and for the ten most diverse clusters. The colour of the most diverse cluster (leftmost point) corresponds to the colour of the name of the cluster shown on the left. The name of the cluster indicates the sub-clustering path : for instance, sub-cluster A7-B6\_D10 means that the 10th sub-cluster of step D comes from the sub-clustering of the 6th cluster from step B, itself coming from the sub-clustering of the 7th initial cluster in step A. The table on the bottom rights gives the differences in the global diversity from one step to the previous one.}
\end{figure}

The procedure described in this paper has been designed for the study of
814 404 spectra of galaxies between redshift 1.2 to 4 from the SDSS DR16
(Abdurro'uf et al., {[}12{]}, 2022) and the DESI DR1 (DESI Collaboration
et al., {[}13{]}, 2023) survey. The full study will be presented in
another paper (Chambon and Fraix-Burnet, in preparation). Here, we
illustrate the procedure on a subsample corresponding to one of the 23
redshift bins dividing the full sample (see e.g. Dubois et al., {[}2{]},
2024). This subsample contains 43826 spectra with redshifts from 1.303
to 1.358. We did not sub-cluster clusters with less than 25 members
because Fisher-EM generally does not converge.

The full procedure generated four levels of sub-clustering (Fig. 3). It
appears that in most cases, at each step, the cluster with the highest
diversity (in the concatenated sub-spaces) corresponds to the cluster
with the highest dispersion of the spectra (in the original variable
space). This could a priori be expected, but a criterion based solely on
the intra-cluster dispersion suffers from the same caveats as the
silhouette score (see Sect. 1) and does not provide an objective
stopping rule. The example in Fig. 3 also shows that diversity and
dispersion are not entirely correlated.

Once a new classification is obtained after a sub-clustering step, it
often happens that the next most diverse cluster is a (sub)cluster of
previous steps (Fig. 3). The whole procedure could thus be drawn on a
dendrogram, but it would have been difficult to illustrate the mean
spectra and the dispersion for all clusters. It must be noted that the
leaves of the dendrogram would in this case be the final classification
rather than individual spectra. At the end, the clusters show comparable
dispersion, generally quite low, and different sizes.

\section{DISCUSSION AND CONCLUSION}

Although the Leinster-Cobbold index is a mathematically rigorously
defined concept, two choices must be made. The first one is the
similarity matrix, that we have taken as the general expression (Eq. 6)
characterized by two quantities, the distance and a constant factor.
They both depend on the data and the perspective of the study. In the
present paper, the high-dimensionality of the data put some constraints
on the choice of the distance.

The second choice is the criterion used to stop the sub-clustering
procedure. The concept of diversity profile yields simple and
quantitative tests, but may need some adjustments. For instance, the
increase of 30\% in the order 0 of the diversity index that we have
chosen, is a compromise between going deep into subsequent
sub-clustering levels (to detect smaller clusters), and finding
significant distinct sub-clusters (to avoid overfitting). This criterion
alone is not sufficient as it essentially counts the number of clusters,
and the other orders (especially 1 and 2) complement it by taking into
account somewhat less rare clusters.

We have found that the diversity measure, initially developed for
biodiversity, is indeed very useful in astrophysics. It helps us solve
the problem of clustering a large amount of spectra of galaxies in an
automatic and objective way. In a previous work (Dubois et al. {[}2{]},
2024), the sub-clustering of a given cluster was decided only from the
number of members, in order to maintain the size of a sub-cluster to at
least the number of variables (to avoid the problem of \(n < p\) in
unsupervised classification). This introduced a bias in favour of the
larger clusters that are preferentially sub-clustered. Instead, the new
procedure presented here favours the rarer clusters of spectra that
increase the diversity. In addition, there is no a priori limitation on
the number of sub-clustering levels, only the diversity index decides.

\section*{REFERENCES}

1. Fraix-Burnet, D., Bouveyron, C. and Moultaka, J. (2021): Unsupervised
classification of SDSS galaxy spectra, \textit{Astronomy and Astrophysics}
, 649, A53.

2. Dubois, J., Siudek, M., Fraix-Burnet, D. and Moultaka, J. (2024)~:
From VIPERS to SDSS: Unveiling galaxy spectra evolution over 9 Gyr
through unsupervised machine-learning, \textit{Astronomy and
Astrophysics,} 687, A76.

3. Zappia, L. and Oshlack, A. (2018): Clustering trees: a visualization
for evaluating clusterings at multiple resolutions, \textit{GigaScience},
7 (7), giy083.

4. Leary, J.R., Xu, Y., Morrison, A.B., Jin, C., Shen, E.C., Kuhlers,
P.C., Su, Y., Rashid, N.U., Yeh, Jen, J. and Peng, X.L. (2023):
Sub-Cluster Identification through Semi-Supervised Optimization of
Rare-Cell Silhouettes (SCISSORS) in single-cell RNA-sequencing,
\textit{Bioinformatics}, 39 (8), btad449.

5. Leinster, T. and Cobbold, C.A. (2012): Measuring diversity: the
importance of species similarity, \textit{Ecology}, 93 (3), 477--489.

6. Bouveyron, C. and Brunet, C. (2012): Simultaneous model-based
clustering and visualization in the Fisher discriminative subspace,
\textit{Statistics and Computing} , 22 (1),

301-324.

7. Hill, M.O. (1973): Diversity and Evenness: A Unifying Notation and
Its Consequences, 54 (2), 427-432.

8. Leinster, T. (2021): \textit{Entropy and Diversity:The Axiomatic
Approach}, Cambridge University Press.

9. Francois, D., Wertz, V. and Verleysen, M. (2007)~: The Concentration
of Fractional Distances, \textit{IEEE Transactions on Knowledge and Data
Engineering}, 19 (7), 873-886.

10. Kabán, A. (2012)~: Non-parametric detection of meaningless distances
in high dimensional data, \textit{Stat Comput}, \textbf{22}, 375--385.

11. Chambon, H. and Fraix-Burnet, D. (2024): Spectral similarities in
galaxies through an unsupervised classification of spaxels,
\textit{Astronomy \& Astrophysics}, 688, A19.

12. Abdurro\textquotesingle uf et al. (2022)~: The Seventeenth Data
Release of the Sloan Digital Sky Surveys: Complete Release of MaNGA,
MaStar, and APOGEE-2 Data, \textit{The Astrophysical Journal Supplement
Series}, 259, 35.

13. DESI Collaboration et al. (2024): The Early Data Release of the Dark
Energy Spectroscopic Instrument, \textit{The Astronomical Journal}, 168,
58.

\end{document}